# A critical assessment of the National Entrepreneurship Context Index of the Global Entrepreneurship Monitor

**Cornelius A. Rietveld***
Erasmus School of Economics
Erasmus University Rotterdam
Burgemeester Oudlaan 50
3062 PA, Rotterdam, The Netherlands
Tel: (010) 408 – 1401
Email: nrietveld@ese.eur.nl
ORCID: https://orcid.org/0000-0003-4053-1861

**Pankaj C. Patel**
Villanova School of Business
Villanova University
800 E. Lancaster Avenue
Villanova, PA 19085, USA
Tel: (610) 519 – 4317
Email: pankaj.patel@villanova.edu
ORCID: https://orcid.org/0000-0003-4934-8166

*Corresponding author

**Abstract**: Data collected through the National Expert Survey (NES) of the Global Entrepreneurship Monitor (GEM) are widely used to assess the quality and impact of national entrepreneurial ecosystems. By focusing on the measurement of the National Entrepreneurship Context Index (NECI), we argue and show that the subjective nature of the responses of the national experts precludes meaningful cross-country analyses and cross-country rankings. Moreover, we show that the limited precision of the NECI severely constraints the longitudinal assessment of within-country trends. We provide recommendations for the current use of NECI data and suggestions for future NES data collections.

**Acknowledgments:** The authors thank Paul Reynolds for his constructive feedback on the rationale and results of this study.

# A critical assessment of the National Entrepreneurship Context Index of the Global Entrepreneurship Monitor

**Abstract**

Data collected through the National Expert Survey (NES) of the Global Entrepreneurship Monitor (GEM) are widely used to assess the quality and impact of national entrepreneurial ecosystems. By focusing on the measurement of the National Entrepreneurship Context Index (NECI), we argue and show that the subjective nature of the responses of the national experts precludes meaningful cross-country analyses and cross-country rankings. Moreover, we show that the limited precision of the NECI severely constraints the longitudinal assessment of within-country trends. We provide recommendations for the current use of NECI data and suggestions for future NES data collections.

## 1. Introduction

Several organizations release country-level indices of institutional conditions for entrepreneurship. Of prominence in entrepreneurship research are the indices and index-based rankings released by the Global Entrepreneurship Monitor (GEM) and the World Bank (WB).[1] Entrepreneurship researchers employ such measures in empirical models to analyze, for instance, the impact of the institutional context and the availability of local resources on new business activity, innovation, or business sentiment (De Clercq, Lim & Oh, 2013; De Jong, Rietveld & Van Stel, 2020), institutional drivers of gender inequality in entrepreneurship (Hechavarría, Terjesen, Stenholm,

---

[1] Less frequently used indices include (in alphabetical order) the Economic Policy Uncertainty Index (Baker, Bloom & Davis, 2016), the Financial Inclusion Index (Cámara & Tuesta, 2014), the Global Entrepreneurship & Development Index (Acs, Szerb & Autio, 2017), the Global Innovation Index (Dutta, Lanvin & Wunsch-Vincent, 2015), the Kauffman Entrepreneurial Activity Index (Fairlie, 2005), the Small Business Sentiment Index (Dunkelberg, 2009), the Startup Meter (http://startup-meter.org/#/), and the World Management Survey (Scur, Sadun, Van Reenen, Lemos & Bloom, 2021).

Brännback & Lång, 2018; Thébaud, 2015), or how business dynamics influence hiring intentions (Haltiwanger, Hyatt, McEntarfer & Sousa, 2012).

The use of these indices of entrepreneurship conditions – and the rankings based on them – is not limited to academic research. Financial institutions incorporate these indices in models quantifying country risk, and money and investments are allocated on the back of such models (Arnold & Jones, 2021; Høyland, Moene, & Willumsen, 2012). Policymakers consider these rankings and year-to-year changes in them as important inputs for their policy- and decision-making (Amorós, Bosma, & Levie, 2013; Coote, 2011). For example, in the introduction of the GEM Jordan National Report (Jordan Enterprise Development Corporation, 2020, p. 2), the Jordanian Minister of Industry, Trade, and Supply states: "Today, the government of Jordan recognizes the importance of the national (GEM) report as a leading global study aimed at improving the understanding of entrepreneurship and promoting evidence-based policy. Jordan in year 2019, had achieved the 34$^{th}$ rank globally in the early stage entrepreneurship activity index, about (9.1%), while this index reached about (8.2%) and ranked 46$^{th}$ globally in year 2017." More problematically, some policymakers pursue high rankings out of prestige: In September 2021, the World Bank Group decided to discontinue the well-known Doing Business reports, as a result of the suspicion of direct and indirect pressure to change the report's methodology to boost China's ranking (World Bank, 2021).

In this *methodological note*, we assess the validity of using data from the National Expert Survey (NES) of GEM to analyse the development of conditions for entrepreneurship across and within countries. Using cross-sectional and longitudinal data, we provide evidence that the subjective nature of the responses of the national experts currently precludes meaningful cross-country analyses and cross-country rankings. That is, while the internal consistency of the National

44Entrepreneurship Context Index measure is high its interrater reliability is poor leading to imprecise country-level measurements. By applying a recently developed methodology (Mogstad, Romano Shaikh & Wilhelm, 2022) to the NECI data, we find that confidence sets for individual country ranks are very wide, and therefore, largely uninformative in separating countries by rankings. For example, in our analysis sample of 54 countries, the confidence set for the highest-ranking country (Indonesia) ranges from rank 1 to rank 27 while it ranges from rank 43 to rank 54 for the lowest-ranking country (Mozambique). Moreover, we show that the imprecision of NECI severely constrains the longitudinal assessment of within-country trends: NECI scores in a particular year cannot be statistically distinguished from the NECI score in the subsequent year.. We conclude by providing recommendations for the current use of NECI data and suggestions for future NES data collections.

## 2. Data

In our study, we use data from the National Expert Survey of the Global Entrepreneurship Monitor (GEM Consortium, 2020b). We use the most recent publicly available NES data as our main data source; these data have been gathered in 2018.[2] In our longitudinal analyses (Section 3.3.2), we use NES data from 2007 to 2018.[3]

Every year, each national team of GEM aims to identify at least 36 experts who rate 9 Entrepreneurial Framework Conditions (EFCs) in their country (at least 4 experts per EFC). Three of these EFCs have two subcomponents, resulting in 12 indicators in total. An overview of these 45conditions is available in Table A1 in the Appendix. In total, the national experts respond to 54 questions regarding the sufficiency of the 12 EFCs in their country using a 9-point Likert scale ranging from "Completely false" (1) to "Completely true" (9).[4] The first item, for instance, reads "In my country, there is sufficient equity funding available for new and growing firms". So, the respondents are asked to evaluate the *sufficiency* of a condition for entrepreneurship *within a particular country*.

---

[2] These data are publicly available via https://www.gemconsortium.org/data/sets?id=nes. Similar analyses as in Sections 3.1, 3.2, and 3.3.1 using earlier releases of GEM's NES data (results available upon request from the authors) show that our conclusions are not driven by the choice to use the 2018 data in the analyses presented in the main text: we observe the same patterns in earlier years.
[3] While GEM data collection started in 1999, individual level NES data are not publicly available for the years 2003, 2004, and 2006. For this reason, we start the longitudinal analyses in 2007.

GEM calculates an expert's score for a particular EFC as the arithmetic mean of the responses given to the relevant items. More recently, in the 2020/2021 report, GEM introduced the National Entrepreneurship Context Index (NECI) as the arithmetic mean of all 54 responses (GEM Consortium, 2020a). This is a convenient way to summarize conditions for entrepreneurship in a country in a single index, and we use this straightforward procedure to construct NECI in the earlier NES data we are analysing here. Table 1 contains the descriptive statistics of the 2018 analysis sample. In total, there are 54 countries in the sample. Due to missing responses on one or more items, there are 1,195 observations for the NECI measure. Thus, on average there are approximately 22 responses from experts per country that can be used to construct NECI. In Table 1, the country ranking reflects the mean score for NECI.

---

[4] Before 2015, responses were collected using a similar 5-point Likert scale for 52 items only (2 items less for Entrepreneurial Finance). In the longitudinal analyses (Section 3.3.2), we therefore use the remapping of the 9-point Likert scale responses to the 5-point Likert scale for the years 2015-2018, as provided by GEM.

**Table 1.** Descriptive statistics and ranking of the 54 countries in the 2018 analysis sample based on the National Entrepreneurship Context Index (NECI).

| Rank | Country | $N$ | NECI (Mean) | NECI (S.D.) | NECI (S.E.) |
|---|---|---|---|---|---|
| 1 | Indonesia | 24 | 6.123 | 1.497 | 0.306 |
| 2 | Canada | 3 | 6.105 | 0.352 | 0.203 |
| 3 | Qatar | 25 | 6.040 | 1.073 | 0.215 |
| 4 | Netherlands | 15 | 5.959 | 0.946 | 0.244 |
| 5 | Taiwan | 25 | 5.856 | 1.301 | 0.260 |
| 6 | India | 36 | 5.724 | 1.421 | 0.237 |
| 7 | USA | 17 | 5.368 | 1.396 | 0.339 |
| 8 | France | 17 | 5.322 | 0.903 | 0.219 |
| 9 | United Kingdom | 7 | 5.217 | 0.824 | 0.311 |
| 10 | Ireland | 17 | 5.041 | 1.185 | 0.287 |
| 11 | Spain | 25 | 4.999 | 0.998 | 0.200 |
| 12 | Luxembourg | 10 | 4.996 | 1.610 | 0.509 |
| 13 | Latvia | 14 | 4.971 | 1.086 | 0.290 |
| 14 | Austria | 19 | 4.937 | 0.915 | 0.210 |
| 15 | United Arab Emirates | 12 | 4.887 | 1.694 | 0.489 |
| 16 | Thailand | 16 | 4.859 | 1.497 | 0.374 |
| 17 | South Korea | 79 | 4.846 | 0.954 | 0.107 |
| 18 | Slovenia | 22 | 4.799 | 0.806 | 0.172 |
| 19 | China (PRC) | 28 | 4.798 | 0.802 | 0.152 |
| 20 | Israel | 17 | 4.788 | 1.025 | 0.248 |
| 21 | Cyprus | 28 | 4.772 | 0.982 | 0.186 |
| 22 | Poland | 22 | 4.769 | 0.703 | 0.150 |
| 23 | Switzerland | 15 | 4.763 | 1.284 | 0.332 |
| 24 | Japan | 30 | 4.725 | 0.873 | 0.159 |
| 25 | Sweden | 11 | 4.677 | 0.778 | 0.235 |
| 26 | Germany | 24 | 4.670 | 1.039 | 0.212 |
| 27 | Turkey | 32 | 4.670 | 1.147 | 0.203 |
| 28 | Mexico | 25 | 4.646 | 1.157 | 0.231 |
| 29 | Chile | 24 | 4.640 | 0.997 | 0.203 |
| 30 | Argentina | 24 | 4.627 | 0.841 | 0.172 |
| 31 | Bulgaria | 21 | 4.329 | 1.182 | 0.258 |
| 32 | Kazakhstan | 20 | 4.320 | 1.279 | 0.286 |
| 33 | Greece | 23 | 4.292 | 1.107 | 0.231 |
| 34 | Colombia | 30 | 4.280 | 1.120 | 0.205 |
| 35 | Lebanon | 18 | 4.235 | 0.750 | 0.177 |
| 36 | Uruguay | 18 | 4.188 | 1.096 | 0.258 |
| 37 | Egypt | 27 | 4.187 | 1.125 | 0.216 |
| 38 | Slovak Republic | 21 | 4.129 | 0.693 | 0.151 |
| 39 | Italy | 27 | 4.060 | 1.035 | 0.199 |
| 40 | Dominican Republic | 27 | 3.872 | 0.694 | 0.134 |
| 41 | Peru | 22 | 3.868 | 1.045 | 0.223 |
| 42 | Brazil | 24 | 3.847 | 1.102 | 0.225 |
| 43 | Morocco | 34 | 3.827 | 0.887 | 0.152 |
| 44 | Guatemala | 23 | 3.799 | 1.070 | 0.223 |
| 45 | Saudi Arabia | 24 | 3.792 | 0.726 | 0.148 |
| 46 | Iran | 36 | 3.771 | 0.921 | 0.153 |
| 47 | Russia | 19 | 3.758 | 1.095 | 0.251 |
| 48 | Panama | 21 | 3.637 | 0.842 | 0.184 |
| 49 | Puerto Rico | 20 | 3.578 | 0.959 | 0.215 |
| 50 | Sudan | 13 | 3.575 | 0.942 | 0.261 |
| 51 | Madagascar | 24 | 3.501 | 0.741 | 0.151 |
| 52 | Croatia | 22 | 3.443 | 0.942 | 0.201 |
| 53 | Angola | 13 | 3.269 | 0.979 | 0.271 |
| 54 | Mozambique | 5 | 2.541 | 0.611 | 0.273 |
| | *Total* | *1,195* | *4.533* | *1.239* | *0.036* |

*Notes*: $N$ denotes the number of experts assessing the national entrepreneurship context in a particular country; S.D.=Standard deviation; S.E.=Standard error; Figure 2 provides a graphical representation of Table 1.
6

## 3. Empirical results

To assess the National Entrepreneurship Context Index (NECI) measure, we look at its internal consistency (Section 3.1), its interrater reliability (Section 3.2), and its validity and precision (Section 3.3) both from a cross-country and longitudinal (within-country) perspective.

### 3.1. Internal consistency

NECI is based on responses of experts that indicate the sufficiency or otherwise of each Entrepreneurial Framework Condition in a particular country. Although experts may have different opinions about what is sufficient or not, even within a country, all the items of NECI are aimed to reflect the same underlying construct. For this reason, we first look at the internal consistency of NECI, which is the consistency of experts' responses across the 54 items. The experts' scores on the items should be correlated with each other, and this can be assessed by estimating Cronbach's alpha. In the 2018 data, Cronbach's alpha for NECI is 0.96, indicating high internal reliability of the measure. Moreover, Cronbach's alpha for the nine main EFCs ranges between 0.74 and 0.92, suggesting that the internal reliability of the underlying EFCs is also high. With GEM (GEM Consortium, 2020a) we, therefore, conclude that the experts responding to the NES rate the items internally consistent.

### 3.2. Interrater reliability

The construction of NECI involves the aggregation of responses. These responses require judgment on the part of the experts. Interrater reliability is the extent to which different experts are consistent in their judgments. Some earlier studies have already suggested that cognitive divergence among the various types (e.g., entrepreneurs, policymakers, and investors) of experts may lead to different ratings (Correia, Silva, Lopes & Braga, 2016; Lee & Wong, 2006; Pfeifer,

Singer, Šarlija & Peterka, 2021). Using a linear regression controlling for country fixed effects, we also find that in the 2018 data the NECI scores of entrepreneurs are significantly lower (0.31) and those of policymakers are significantly higher (0.17) than the average score.

Given the continuous nature of NECI, we computed the interrater reliability as the intraclass correlation coefficient (ICC) from a one-way random-effects model. Strikingly, the ICC is as low as 0.30 showing that interrater reliability is poor. To verify that this estimate is not driven by the lower number of experts in some countries, we duplicated the data ten times. In this expanded data set, the interrater reliability is also poor (0.33), suggesting that the main estimate mostly reflects disagreements across raters. When we focus on the country with the highest number of responses, South Korea, it is evident that evaluations indeed vary greatly among experts. While the mean rating is 4.85, the minimum and maximum ratings are respectively 2.46 and 7.96 (Figure 1). We, therefore, conclude that NECI, despite its internal consistency (Section 3.1), is not reliably measured.

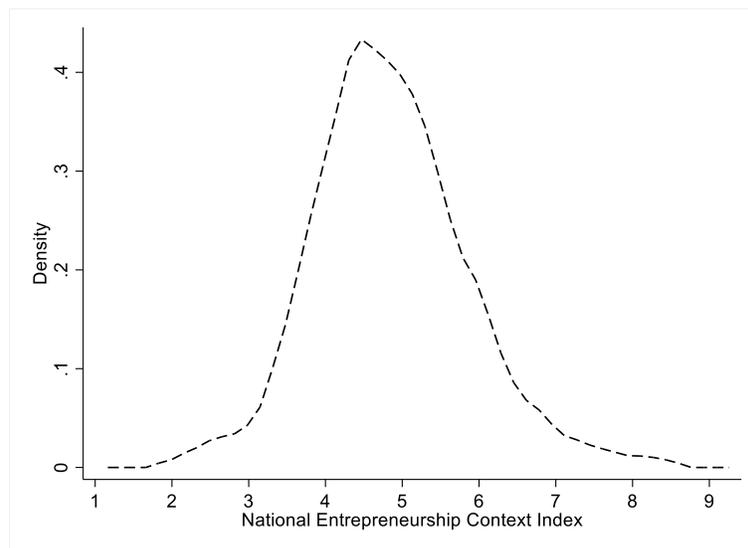

**Figure 1.** Kernel density plots of the distribution of the scores given by experts assessing the 2018 national entrepreneurship context in South Korea (79 experts).

## 3.3. The precision of NECI

*3.3.1. Cross-sectional precision*

The dispersion of the expert ratings (Section 3.2) leads to imprecise NECI scores at country-level, which is evident from the standard deviation and standard error of the country-level means. For example, Table 1 shows that for the country with the highest rank (1), Indonesia, the mean score of 6.123 is based on the responses of 24 experts. As a result, the standard error of the mean score is relatively large, i.e., 0.306. Thus, the mean NECI scores of the countries ranking 2-5 (Canada, Qatar, the Netherlands, and Taiwan) are within 1 standard error from the mean NECI score of Indonesia. Several studies adopt these continuous indices as predictor variables as if they come without uncertainty (e.g., De Clercq, Lim & Oh, 2013; Hechavarría, Terjesen, Stenholm, Brännback & Lång, 2018). Figure 2, as visualization of Table 1, shows that this is unwarranted given that the mean scores are considerably imprecise.

**Figure 2.** Mean score (with 95% confidence interval) of the National Entrepreneurship Context Index for the 54 countries in the 2018 analysis sample.

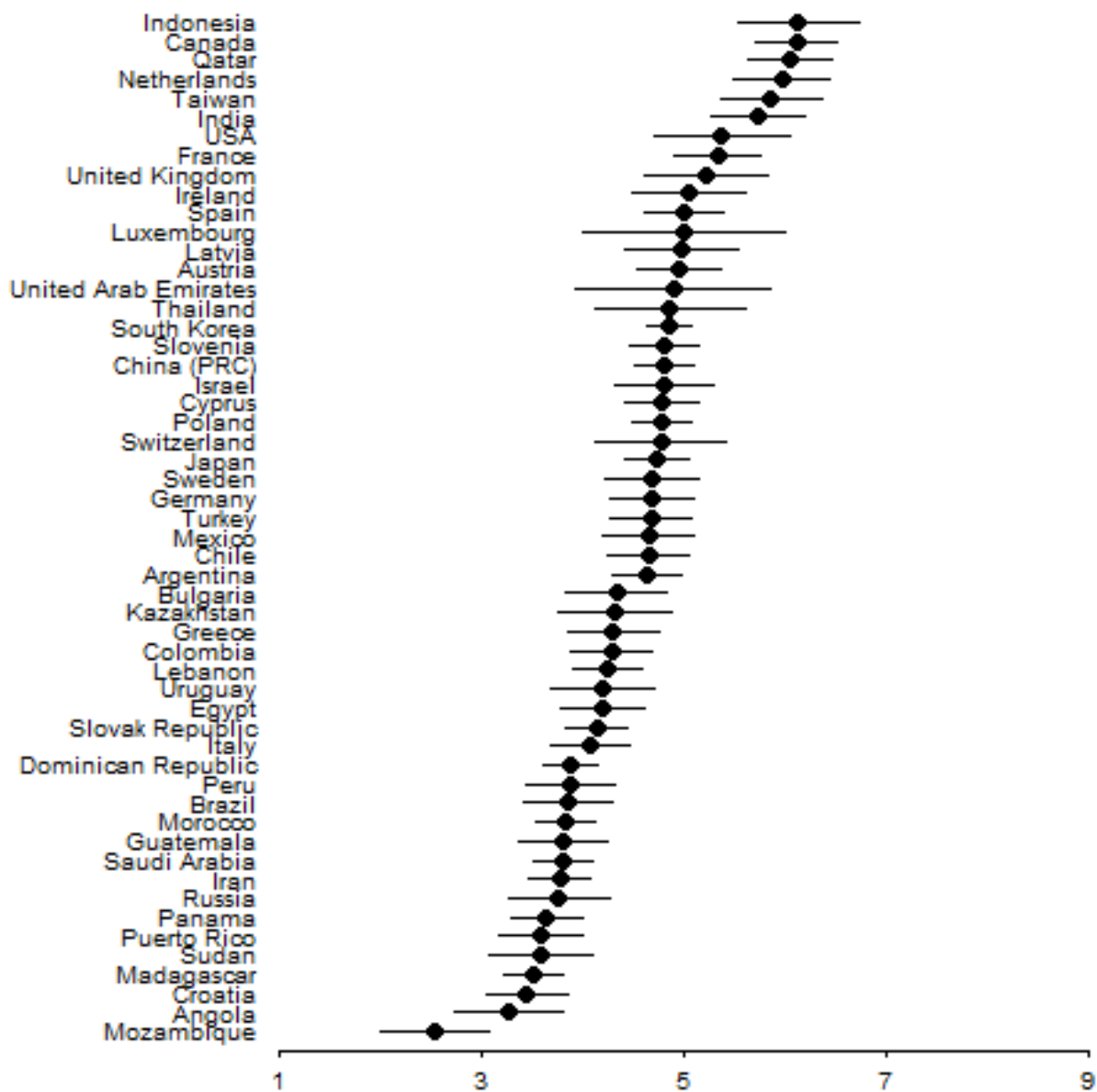

In Table 1, we ranked countries based on their mean score for NECI. Such a ranking is an important part of the yearly GEM reports, but ignores the imprecision of the mean scores. Recently, Mogstad et al. (2022) developed a methodology to derive confidence sets for rankings based on mean scores that does take into account the uncertainty surrounding these means. The goal of this methodology is to construct confidence sets that contain the ranks of all countries with a probability approximately not less than some pre-specified level (e.g., 95%). To construct simultaneous confidence sets for the ranking of all 54 countries, one first needs to compute the difference in mean NECI estimates for all pairwise combinations of countries. The confidence set for a country (e.g., Indonesia) is the product of these (54 minus 1=) 53 differences such that the probability of it simultaneously covering all differences is at least 95%. The bounds for the simultaneous confidence sets depend on quantiles from the distribution over the maximum of the estimated differences and can be obtained using bootstrap.[5] Then, given the simultaneous confidence sets, the ranking is determined by how many confidence sets for a particular country lie above and below zero (because a difference of 0 would indicate no difference in mean score). For a detailed description of the methodology, we refer readers to Mogstad et al. (2022).

Figure 3 visualizes the results of this procedure for the countries in the NES data of GEM 2018. While Indonesia ranks 1, its confidence set indicates that Indonesia ranks with 95% certainty between ranks 1 and 27. In other words, due to uncertainty in the aggregation of the responses of experts, there is 95% uncertainty that Indonesia ranks between 1 and 27 in the set of 54 countries. When we look at the bottom of the ranking, we see that the rank of Mozambique is ranks between 43 and 54. While this is clearly at the lower end of the distribution, Mozambique is only 1 of the 12 countries with a rank of 54 in its confidence set. These estimates provide an important inference: Wide variation in the precision of individual rankings renders country rankings less meaningful.[6]

---

[5] The quantiles depend on the choice of the pre-specified confidence level; Analysis code (in *R*) to run this statistical procedure is freely available via https://github.com/danielwilhelm/R-CS-ranks.

[6] Similar plots were created for each of the twelve individual EFCs (results available upon request from the authors). As a result of larger sample sizes for some of the EFCs, the confidence sets for some EFCs are a bit narrower than the confidence sets for other EFCs. Still, the same conclusion holds as for the NECI analysis: Rankings are very uninformative, with confidence sets spanning almost all ranks for a large number of countries.

**Figure 3.** Ranking of the 54 countries participating in the Global Entrepreneurship Monitor 2018 based on the National Entrepreneurship Context Index (with 95% confidence sets).

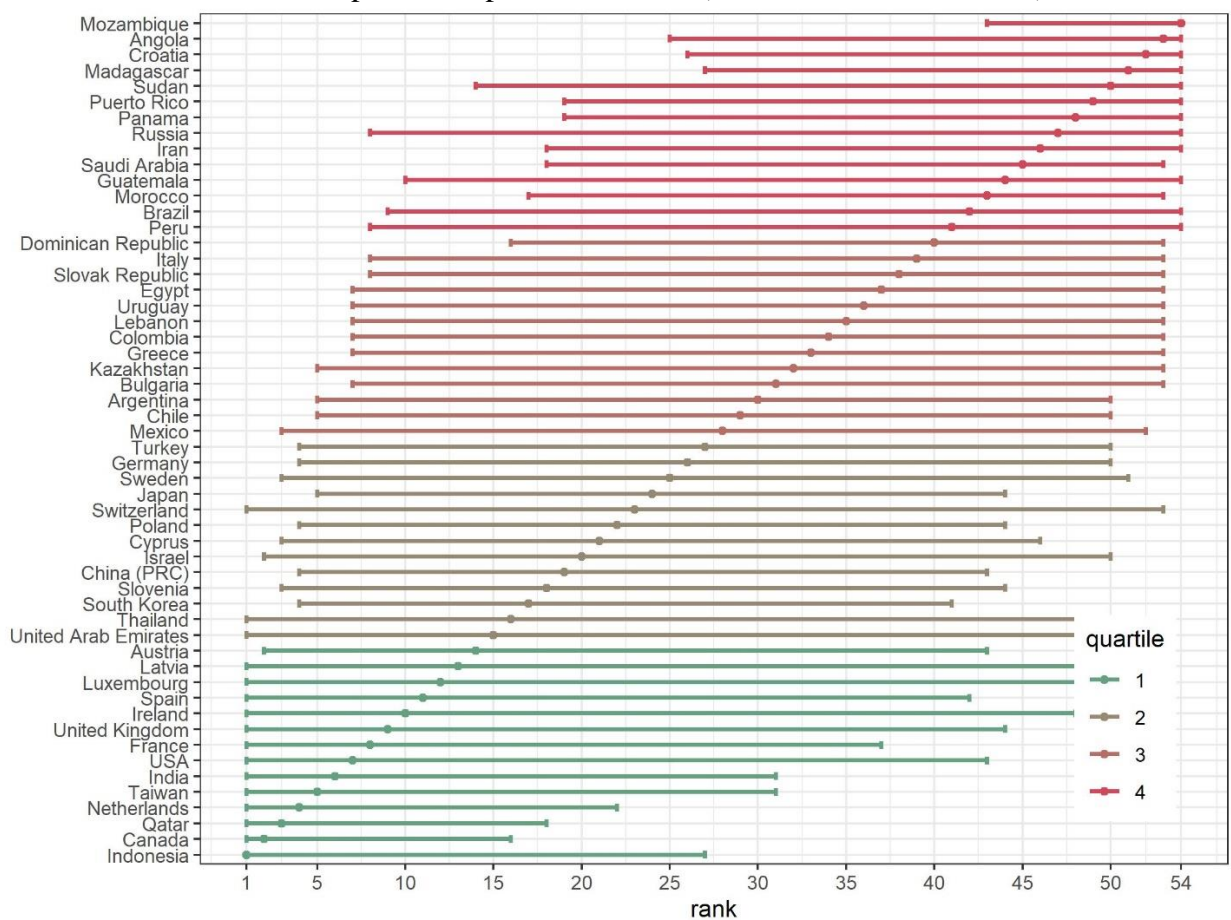

*3.3.2. Longitudinal precision*

GEM acknowledges that the subjective nature of the NES may hamper cross-country analyses, because "National experts in a particular developed economy, for example, may have a very different view of sufficiency from national experts in a much less-developed economy" (Bosma et al., 2021, p. 77). While we have seen that the poor reliability of the NECI (Section 3.2) leads to imprecise country-level NECI scores (Section 3.3.1), it is, however, difficult to verify the cross-country validity of NECI. That is, how accurately NECI measures the overall conditions for entrepreneurship in a country, because there is no perfect measure of the underlying construct available. Nevertheless, we find that the correlation between the NECI score and the Doing

Business score of the World Bank is only 0.56 ($p < 0.001$) in 2018. While both measures may not perfectly capture conditions for entrepreneurship in a country, this moderate correlation is likely to be partially driven by the subjective nature of the NES data. Therefore, we here analyze whether the development of NECI scores *within* a country may still be informative. A longitudinal analysis within a single country is not being hampered by different view of sufficiency across countries. Figure 4 shows the development of NECI in the seven countries that have complete information on it for the years 2007 until 2018.

**Figure 4.** The development of the National Entrepreneurship Context Index (5-point scale, with 95% confidence intervals) in the period 2007-2018 for the seven countries with complete information (the dashed line represent the mean NECI score in the analysed period).

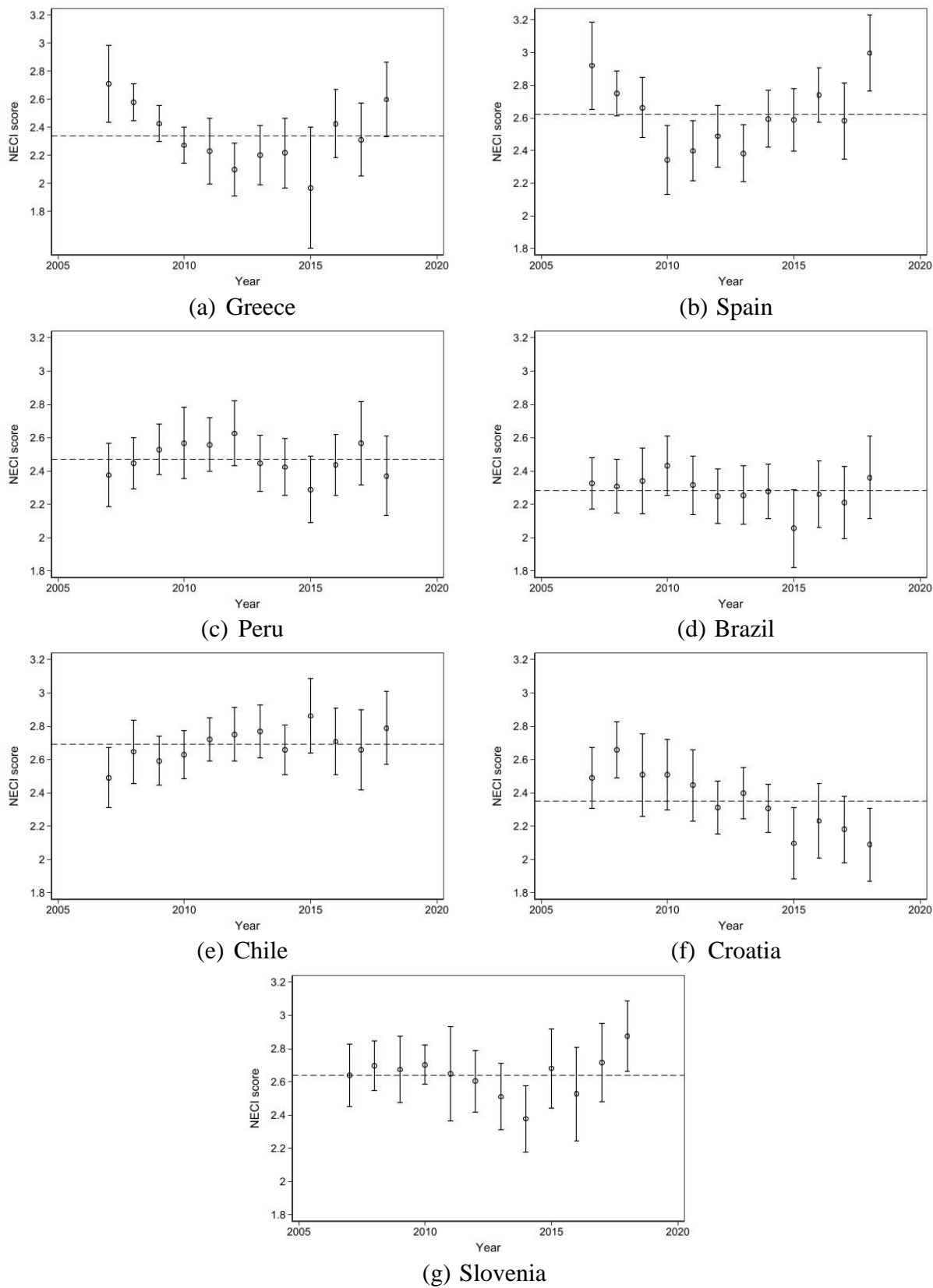

(a) Greece  (b) Spain  (c) Peru  (d) Brazil  (e) Chile  (f) Croatia  (g) Slovenia

15151515151515151515151515

It is noteworthy that from the (7×12=) 84 data points in Figure 4, only 12 are significantly different from the respective country mean at the 95% confidence level. For these 12 data points, the deviation from the country mean is usually very small. For Peru, Brazil, and Chile, the NECI scores in the period between 2007 and 2018 cannot be distinguished from their country mean in any of the years. Importantly, in none of the countries the NECI score in a particular year can be statistically distinguished from the NECI score in the subsequent year, clearly showing that there is limited value in interpreting year-to-year changes in the NECI scores when properly accounting for uncertainty.

## 4. Discussion

Several indices are used to measure conditions for entrepreneurship in countries, regions, or states. Among them, the National Entrepreneurship Context Index as developed by the Global Entrepreneurship Monitor is of prime importance for researchers and policy makers. While the internal consistency of this measure is high, the subjective nature of the responses of the national experts results in poor interrater reliability. Experts may have diverging views of what "sufficiency" means, even within a country. Therefore, our first recommendation is that GEM adopts vignettes (Alexander & Becker, 1978) in future NES surveys to calibrate beliefs about "sufficiency". By doing so, the interrater reliability of the NES data may improve. Relatedly, to measure NECI, GEM draws on responses of at least 36 experts in a country. Table 1 shows that due to missingness the NECI scores are on average computed based on fewer responses. Given the costs of data collection, reducing item missingness to the minimum would be key in future NES data collections.

Establishing the validity of NECI as a measure of the conditions for entrepreneurship is not straightforward, but it is clear that NECI already comes with such imprecision that meaningful cross-country analyses and cross-country rankings are difficult to achieve. Country-level studies currently use the country-level NES index data as if they come without imprecision (e.g., De Clercq, Lim & Oh, 2013; Hechavarría, Terjesen, Stenholm, Brännback & Lång, 2018). When the imprecision (measurement error) of the variable of interest is correlated with the true value of that variable, with the true values of other variables in the model, or with the errors in measuring those values, bias in the estimation of the coefficient can be both downward or upward depending on the association between the measurement error in the variable of interest variables and the other variables included in the model (Bound, Brown & Mathiowetz, 2001; Hausman, 2001). In the context of NES data, this is the likely scenario because the number of experts assessing EFCs differs across countries and "sufficiency" may mean different things in different countries. For this reason, it does not seem justified to continue with regression-based cross-country analyses using the currently available NES data.

Rankings are an important part of the yearly GEM reports, but when not taking into account the uncertainty in these rankings one runs the risk of overinterpreting a particular rank. This has been coined before "tyranny of international index rankings" (Høyland, Moene & Willumsen, 2012), and it may lead to misallocation of resources. For example, high rankings may suggest efficacious country-specific resources and activities around entrepreneurship, and therefore, policymakers in higher-ranked countries may expect high returns from investments in entrepreneurship programs. As a consequence, policymakers in countries or regions in the upper ranks may decide to turn their attention to other high-priority non-entrepreneurship areas. When properly accounting for ranking uncertainty using the methodology recently developed by Mogstad et al. (2002), the ranking of countries based on the National Expert Survey of the Global Entrepreneurship Monitor appears to be largely illusory: The confidence sets are wide and tend to span many ranks for each specific country. As such, year-to-year changes in the ranking of a particular country based on NES data cannot be interpreted as structural changes.[7]

It is, however, a great advantage that all individual-level responses of the national experts are made publicly available by GEM, making it possible to directly apply Mogstad et al. (2022)'s methodology to these data. While the Global Entrepreneurship & Development Index and World Bank Doing Business rankings are also often used in entrepreneurship research, they are based on weighted inputs from different sources making it more difficult to apply the Mogstad et al. (2022) methodology to these indices because the calculation of the standard errors is not straightforward. The developers of GEDI appreciate the uncertainty in the rankings they provide although they are not fully able to assess this uncertainty (i.e., to provide the correct standard error). Høyland, Moene, and Willumsen (2012) developed a procedure to estimate credibility intervals for the World Bank Doing Business scores and go on to show that these scores are highly imprecise. Our results for GEM's NES data align with these results, although the latter study does not derive confidence sets for the ranking of individual countries (and the credibility intervals cannot be directly used as input for the Mogstad et al. (2022) methodology).

Therefore, if GEM wants to maintain country rankings in their reports, it is advised to adopt Mogstad et al. (2022)'s methodology and to report confidence set for rankings rather than point estimates to lower overfocus on the absolute levels of rankings. The confidence sets may reduce the plausible 'arms race' based on differences in rankings that may not be statistically different. Allowing policymakers to consider the point estimate for rankings and the associated confidence bounds provides a more meaningful set of decision-making criteria and lowers singular focus on a particular ranking. One straightforward way to tighten the confidence sets, besides improving the interrater reliability of NECI, is to increase the number of experts solicited for their evaluations in the NES. However, Figure A1 in the Appendix visualizes that even in the case of a total of 36 responses per country (the current target of GEM) the obtained NECI ranking will be imprecise.[8] Increasing the expert responses to 100 per country may help separate the highest-ranking countries from the lowest-raking countries (Figure A2 in the Appendix), but this will be costly and the rankings of countries in the middle of the distribution will still be considerably imprecise.

---

[7] The set of countries participating in GEM changes every year, something that is also often not taken into account when interpreting year-to-year changes in rankings (*cf.* the Jordanian GEM report).

The subjective nature of the NES scores severely limits the informative ranking of conditions for entrepreneurship *across* countries, but the development of the scores *within* a country over time may nevertheless be informative especially when the interrater reliability improves (for example, through the use of vignettes) and when the number of experts solicited increases. Currently, despite the yearly collection of data, the NECI scores are too imprecise to meaningfully analyze within-country trends. We recommend GEM to transparently release the imprecision of their measures in their communications (for example, by using figures like Figure 2 and Figure 4) as well as in their country-level data releases (for example, by including standard errors of country-level aggregates in the data sets). This will help researchers and policy makers to present and consider within-country longitudinal trends with an adequate level of precision.

## 5. Conclusion

Resources, incentives, markets, and supporting institutions for the creation and growth of new firms are embedded in institutional and cultural contexts that tend to change only slowly 419(Audretsch, Falck, Feldman & Heblich, 2012; Thurik & Dejardin, 2012). As a result, country differences in entrepreneurship are relatively stable over time (Bosma et al., 2021; Freytag & Thurik, 2007; Mathers, 2013). The widespread release of indices and rankings to quantify entrepreneurial activity and conditions for entrepreneurship facilitates straightforward cross-country and longitudinal comparisons for policymakers and the general public. While the NES data of GEM, in particular the National Entrepreneurship Context Index, may add to this, our assessments show that the subjective nature of the responses of the national experts currently precludes meaningful cross-country analyses and cross-country rankings. Moreover, we show that the imprecision of the NECI severely constraints the longitudinal assessment of within-country trends. While the interrater reliability of NECI is currently low, its internal consistency is high. Therefore, we believe that by following the outlined suggestions in the previous section the NECI may realize its potential in future NES data collections.

---

[8] For these analyses, we use the means and standard deviations as reported in Table 1 as input; the standard error of the mean estimate is set equal to the standard deviation divided by the squareroot of the chosen sample size.

22## Appendix

**Table A1.** Entrepreneurial Framework Conditions assessed in the National Experts Survey of the Global Entrepreneurship Monitor.

| # | Entrepreneurial Framework Condition | Description |
|---|---|---|
| 1 | Entrepreneurial Finance (8 questions in 2015-2018; 6 question in 2007-2014) | The availability of financial resources—equity and debt—for small and medium enterprises (SMEs) (including grants and subsidies) |
| 2a | Government Policy (7 questions) | The extent to which public policies support entrepreneurship: a) Entrepreneurship as a relevant economic issue new and SMEs. |
| 2b | | The extent to which public policies support entrepreneurship: b) Taxes or regulations are either size-neutral or encourage |
| 3 | Government Entrepreneurship Programs (6 questions) | The presence and quality of programs directly assisting SMEs at all levels of government (national. regional. municipal) |
| 4a | Entrepreneurship Education (6 questions) | The extent to which training in creating or managing SMEs is incorporated within the education and training system at all levels: a) Entrepreneurship Education at basic school (primary and secondary) |
| 4b | | The extent to which training in creating or managing SMEs is incorporated within the education and training system at all levels: b) Entrepreneurship Education at post-secondary levels (higher education such as vocational. college. business schools. etc.) |
| 5 | R&D Transfer (6 questions) | The extent to which national research and development will lead to new commercial opportunities and is available to SMEs |
| 6 | Commercial and Legal Infrastructure (5 questions) | The presence of property rights. commercial. accounting. and other legal and assessment services and institutions that support or promote SMEs. |
| 7a | Entry Regulation (6 questions) | a) Market Dynamics: the level of change in markets from year to year |
| 7b | | b) Market Openness: the extent to which new firms are free to enter existing markets |
| 8 | Physical Infrastructure (5 questions) | Ease of access to physical resources—communication. utilities. transportation. land. or space—at a price that does not discriminate against SMEs |
| 9 | Cultural and Social Norms (5 questions) | The extent to which social and cultural norms encourage or allow actions leading to new business methods or activities that can potentially increase personal wealth and income |

*Source:* Global Entrepreneurship Monitor. https://www.gemconsortium.org/wiki/1142.

**Figure A1.** Ranking of the 54 countries participating in the Global Entrepreneurship Monitor 2018 based on the National Entrepreneurship Context Index (with 95% confidence sets), assuming the means and standard deviations of expert responses as in Table 1 and $N_{experts} = 36$).

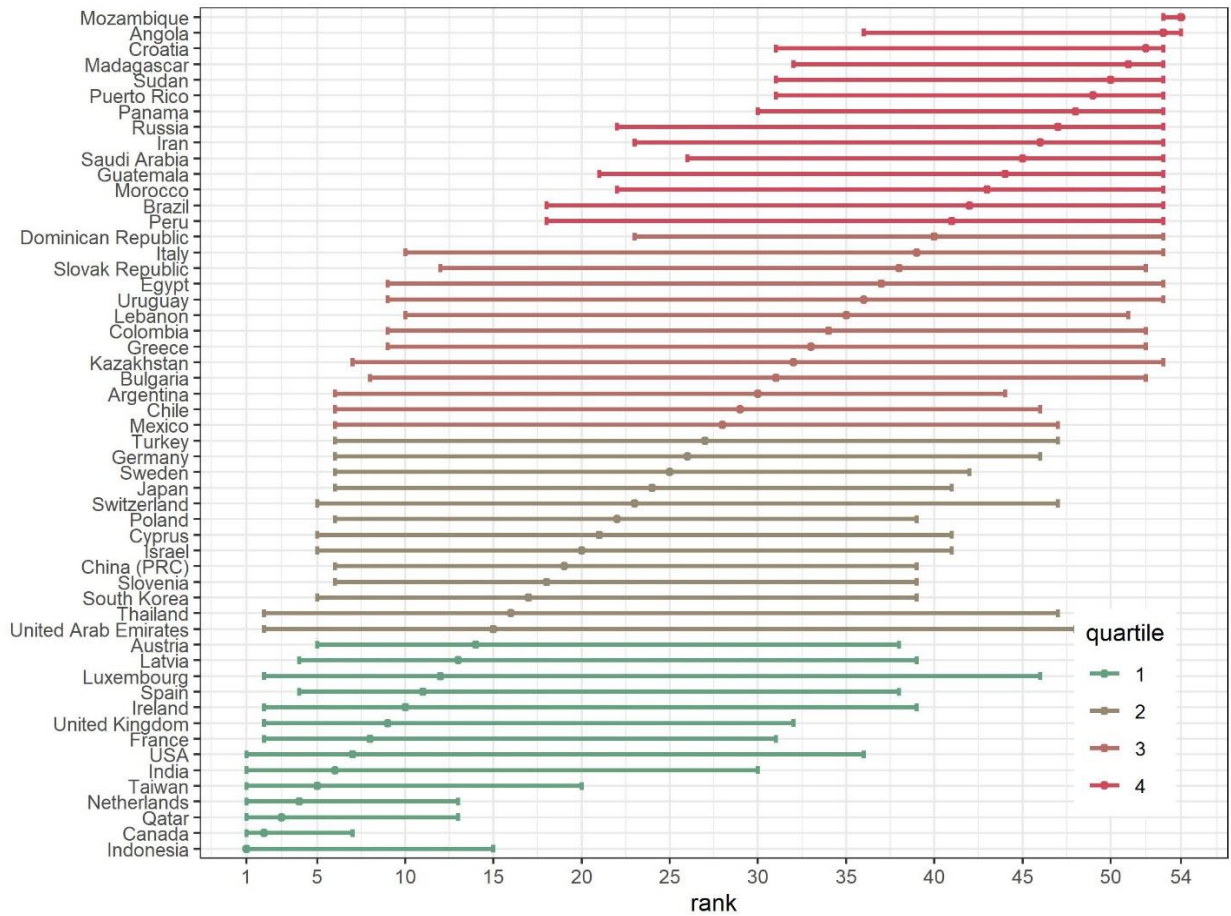

**Figure A2.** Ranking of the 54 countries participating in the Global Entrepreneurship Monitor 2018 based on the National Entrepreneurship Context Index (with 95% confidence sets), assuming the means and standard deviations of expert responses as in Table 1 and $N_{\text{experts}} = 100$).

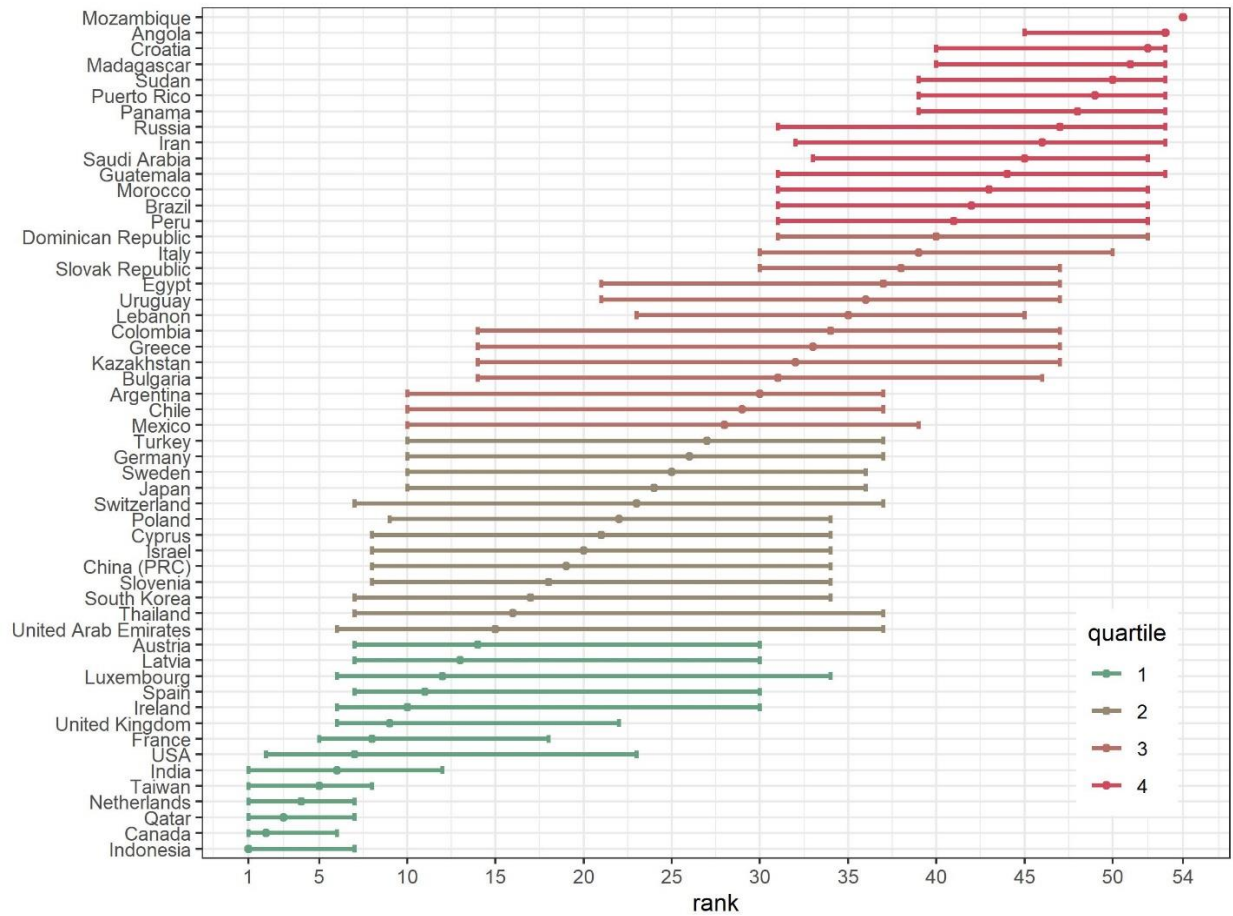